\documentclass[traditabstract]{aa}
 \usepackage[a4paper, left=1.5cm, right=1.5cm, bottom=2.5cm, top=2.5cm]{geometry}
 \usepackage[utf8]{inputenc}
\usepackage{txfonts}
\usepackage{graphicx}
\usepackage{array}
\usepackage{natbib}
\usepackage{multirow}
\usepackage{relsize}
\usepackage{amsbsy}
\usepackage{hyperref}
\usepackage{epstopdf}

\DeclareGraphicsExtensions{.eps}

\begin{document}

\title {
The stellar mass -- halo mass relation from galaxy clustering in VUDS: a high star formation efficiency at z$\simeq$3 \thanks{Based on data obtained with the European Southern Observatory Very Large
Telescope, Paranal, Chile, under Large Program 185.A-0791.}} 
\titlerunning{The stellar-to-halo mass relation at z$\sim3$ in VUDS}

\author{A. Durkalec \inst{1}
\and O. Le F\`evre\inst{1}
\and S. de la Torre\inst{1}
\and A. Pollo\inst{19,20}
\and P. Cassata\inst{1,18}
\and B. Garilli\inst{3}
\and V. Le Brun\inst{1}
\and B.C. Lemaux \inst{1}
\and D. Maccagni\inst{3}
\and L. Pentericci\inst{4}
\and L.A.M. Tasca\inst{1}
\and R. Thomas\inst{1}
\and E. Vanzella\inst{2}
\and G. Zamorani \inst{2}
\and E. Zucca\inst{2}
\and R. Amor\'in\inst{4}
\and S. Bardelli\inst{2}
\and L.P. Cassar\`a\inst{3}
\and M. Castellano\inst{4}
\and A. Cimatti\inst{5}
\and O. Cucciati\inst{5,2}
\and A. Fontana\inst{4}
\and M. Giavalisco\inst{13}
\and A. Grazian\inst{4}
\and N. P. Hathi\inst{1}
\and O. Ilbert\inst{1}
\and S. Paltani\inst{9}
\and B. Ribeiro\inst{1}
\and D. Schaerer\inst{10,8}
\and M. Scodeggio\inst{3}
\and V. Sommariva\inst{5,4}
\and M. Talia\inst{5}
\and L. Tresse\inst{1}
\and D. Vergani\inst{6,2}
\and P. Capak\inst{12}
\and S. Charlot\inst{7}
\and T. Contini\inst{8}
\and J.G. Cuby\inst{1}
\and J. Dunlop\inst{16}
\and S. Fotopoulou\inst{9}
\and A. Koekemoer\inst{17}
\and C. L\'opez-Sanjuan\inst{11}
\and Y. Mellier\inst{7}
\and J. Pforr\inst{1}
\and M. Salvato\inst{14}
\and N. Scoville\inst{12}
\and Y. Taniguchi\inst{15}
\and P.W. Wang\inst{1}
}

\institute{Aix Marseille Universit\'e, CNRS, LAM (Laboratoire d'Astrophysique de Marseille) UMR 7326, 13388, Marseille, France
\and
INAF--Osservatorio Astronomico di Bologna, via Ranzani,1, I-40127, Bologna, Italy
\and
INAF--IASF, via Bassini 15, I-20133, Milano, Italy
\and
INAF--Osservatorio Astronomico di Roma, via di Frascati 33, I-00040, Monte Porzio Catone, Italy
\and
University of Bologna, Department of Physics and Astronomy (DIFA), V.le Berti Pichat, 6/2 - 40127, Bologna, Italy
\and
INAF--IASF Bologna, via Gobetti 101, I--40129, Bologna, Italy
\and
Institut d'Astrophysique de Paris, UMR7095 CNRS,
Universit\'e Pierre et Marie Curie, 98b Boulevard Arago, 
Paris, France
\and
Institut de Recherche en Astrophysique et Plan\'etologie - IRAP, CNRS, Universite de Toulouse, UPS-OMP, 14, avenue E. Belin, F31400
Toulouse, France
\and
Department of Astronomy, University of Geneva,
ch. d'cogia 16, CH-1290 Versoix, Switzerland
\and
Geneva Observatory, University of Geneva, ch. des Maillettes 51, CH-1290 Versoix, Switzerland
\and
Centro de Estudios de F\'isica del Cosmos de Arag\'on, Teruel, Spain
\and
Department of Astronomy, California Institute of Technology, 1200 E. California Blvd., MC 249--17, Pasadena, USA
\and
Astronomy Department, University of Massachusetts, Amherst, MA 01003, USA
\and
Max-Planck-Institut f\"ur Extraterrestrische Physik, Postfach 1312, D-85741, Garching bei M\"unchen, Germany
\and
Research Center for Space and Cosmic Evolution, Ehime University, Bunkyo-cho 2-5, Matsuyama 790-8577, Japan
\and
SUPA, Institute for Astronomy, University of Edinburgh, Royal Observatory, Edinburgh, EH9 3HJ, United Kingdom
\and
Space Telescope Science Institute, 3700 San Martin Drive, Baltimore, MD 21218, USA 
Instituto de Fisica y Astronomia, Facultad de Ciencias, Universidad de Valparaiso, Av. Gran Bretana 1111, Casilla 5030, Valparaiso, Chile
\and
Astronomical Observatory of the Jagiellonian University, Orla 171, 30-001 Cracow, Poland
\and
National Centre for Nuclear Research, ul. Hoza 69, 00-681, Warszawa, Poland
}


\abstract{The relation between the galaxy stellar mass M$_{\star}$ and the dark
  matter halo mass M$_h$ gives important information on the
  efficiency in forming stars and assembling stellar mass in galaxies.
  We present the stellar mass to halo mass ratio (SMHR) measurements at redshifts $2<z<5$,
  obtained from the VIMOS Ultra Deep Survey. 
  We use halo occupation distribution (HOD) modelling of clustering measurements on $\sim$3000
  galaxies with spectroscopic redshifts to derive 
  the dark matter halo mass M$_h$, and SED fitting over a large set of multi-wavelength data
  to derive the stellar mass M$_{\star}$ and compute the SMHR=M$_{\star}$/M$_h$. 
  We find that the SMHR ranges from 1\% to 2.5\% for galaxies with M$_{\star}$=$1.3\times10^9$ $M_{\sun}$ 
  to M$_{\star}$=$7.4\times10^9$ $M_{\sun}$ in DM halos with M$_h$=$1.3\times10^{11}$ M$_{\sun}$ to
  M$_h$=$3\times10^{11}$ M$_{\sun}$.
  We derive the integrated star formation efficiency (ISFE) of these galaxies and find that the star formation
  efficiency is a moderate 6--9\% for lower mass galaxies while it is relatively  high at 16$\%$ for
  galaxies with the median stellar mass of the sample $\sim7\times10^9$ $M_{\sun}$. 
  The lower ISFE at lower masses may indicate that some efficient means of
  suppressing star formation is at work (like SNe feedback), while the high ISFE for the average galaxy at z$\sim$3 is 
  indicating that these galaxies are efficiently building-up their stellar mass at a key epoch in
  the mass assembly process. We further infer that the average mass galaxy at z$\sim$3 will 
  start experiencing star formation quenching within a few hundred millions years.
}


\maketitle

\renewcommand{\arraystretch}{1.5}

\section{Introduction}

Understanding processes regulating star formation and mass growth in galaxies along cosmic time remains a key issue of galaxy formation and evolution.
In the  $\Lambda$CDM model dark matter (DM) halos grow hierarchically, and galaxies are thought to form via dissipative collapse in the deep potential wells of these DM halos (e.g. \citeauthor{White1978} \citeyear{White1978}, \citeauthor{Fall1980} \citeyear{Fall1980}). 
In this paradigm, cooling processes bring baryons in high density peaks of the matter density field (haloes), where the conditions for gas fragmentation trigger star formation \citep{Bromm2009}. 
Current models connecting star formation and stellar mass evolution on the one hand, and the formation histories of DM halos on the other hand, are relying on simplifying assumptions and approximations and need to be further informed by observational data to reduce the uncertainties in the modelling process (e.g. \citeauthor{Conroy2009a} \citeyear{Conroy2009a}).

The efficiency of assembling baryons into stars is an important ingredient to understand galaxy formation but remains poorly constrained observationally. 
In recent years it has been proposed to derive this efficiency comparing DM halo mass with galaxy stellar mass. 
With the measurement of the characteristic mass of DM host haloes M$_h$ now available from observational data and of stellar mass M$_{\star}$ derived  from the analysis of the spectral energy distribution of galaxies, coupled to the  knowledge of the cosmological density of baryons and DM, one can infer the  conversion rate from baryons to stellar mass.

Two methods have been used so far to link M$_{\star}$ and M$_h$. 
Halo occupation models provide a description of how galaxies populate their host haloes using galaxy clustering statistics and local density profiles (e.g. \citeauthor{Zehavi2005} \citeyear{Zehavi2005}, \citeauthor{Leauthaud2012} \citeyear{Leauthaud2012}). 
Alternatively, abundance matching associate galaxies to underlying dark matter structure and sub-structures assuming that the stellar masses or luminosities of the galaxies are tightly connected to the masses of dark matter 
halos (\citeauthor{Conroy2009b} \citeyear{Conroy2009b}, \citeauthor{Moster2013} \citeyear{Moster2013}). 
The efficiency with which the galaxies converted 
baryons into stars is encoded in the relationship between $M_{\star}$ and $M_h$ as a function of redshift, which provides a benchmark against which galaxy evolution models can be tested.
Using observed stellar mass functions, abundance matching models have led to the derivation of the 
Stellar Mass -- Halo Mass (SMHM) relation which gives for a given halo mass the 
Stellar Mass to Halo Mass ratio (SMHR), SMHR$=$M$_{\star}$/M$_h$. 
\cite{Behroozi2010} find that the integrated star formation efficiency (ISFE) at a given halo mass peaks
at 10-20\% of available baryons for all redshifts from 0 to 4.

The shape of the SMHM is claimed not 
to evolve much from $z=0$ to $z=4$, although it may be evolving more significantly at $z>4$ (\citeauthor{Behroozi2013} \citeyear{Behroozi2013}, \citeauthor{Behroozi2014} \citeyear{Behroozi2014}). 
The SMHR is characterized by a maximum around M$_h=$10$^{12}$M$_{\sun}$. 
The lower efficiency at masses below this value may indicate
that supernova feedback might be sufficient to remove gas from the galaxy 
as the halo gravitational potential is lower (e.g. \citeauthor{Silk2003} \citeyear{Silk2003}, \citeauthor{Bertone2005} \citeyear{Bertone2005}, 
\citeauthor{Bethermin2013} \citeyear{Bethermin2013}). 
At higher masses, the decrease in star formation efficiency might be produced when cold streams are replaced
by isotropic cooling (e.g. \citeauthor{Dekel2006} \citeyear{Dekel2006}, \citeauthor{Faucher2011} \citeyear{Faucher2011}) or by some 
high energy feedback process like that produced by AGNs. 

While this picture is attractive from a theoretical modelling point of view, consistency with observational constraints need to be further improved. In this Letter we use the VIMOS Ultra Deep Survey (VUDS, \citeauthor{OLF2014} \citeyear{OLF2014}) to report on
the first measurements of the SMHR derived from the observed clustering of galaxies at $2<z<5$. Using M$_h$ derived from HOD modelling based on the two-point projected correlation function $w_p(r_p)$, and $M_{\star}$ obtained from SED fitting computed from $\sim$3000 galaxies we estimate the SMHR for several galaxy samples, and compare it to SMHM models.
The Letter is organized as follows: we summarize the VUDS data in Section \ref{sec:2_data}, 
the M$_h$ and M$_{\star}$ measurements are presented in Section \ref{sec:3_methods}, we derive the SMHR
and the ISFE for several mass bins at z$\sim$3 in Section \ref{sec:4_results}, and we discuss our results in 
Section \ref{sec:5_discussion}.

We use a flat $\Lambda CDM$ cosmological model with $\Omega_m = 0.25$,
and a Hubble constant $H_0=70$ km $\textrm{s}^{-1}$ Mpc to compute absolute magnitudes and  masses.

\begin{figure}[t]
  \includegraphics[height=\hsize, angle=270]{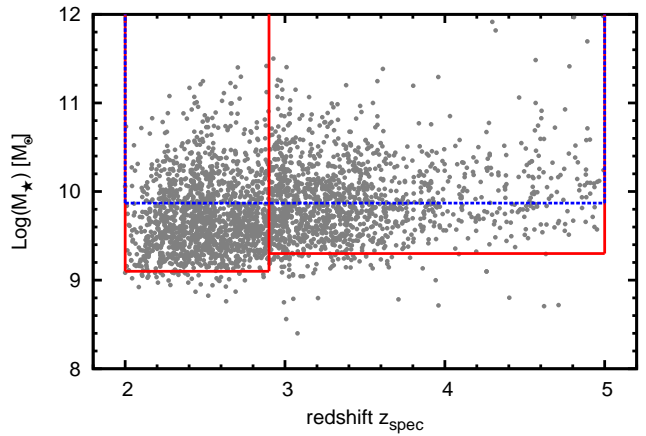}
  \caption{Stellar mass distribution in VUDS. 
	  Red lines and horizontal lines indicate the limits in redshift and stellar mass applied
          to select low and high redshift samples. 
	  The dashed blue line indicate the mass cut at M$_{\star}$=7.4$\times10^9$ M$_{\sun}$ applied to define a high-mass sample.}
  \label{fig:stellar_mass}
\end{figure}

\begin{table*}[t]
  \caption{The stellar mass to halo mass ratio (SMHR), and the integrated star formation efficiency (ISFE) in the VUDS survey}
  \noindent\hrulefill\par
  \noindent\makebox[7cm][c]{%
  \begin{minipage}{5cm}
  \centering
    \scalebox{1.2}{
  \begin{tabular}{lccccccc} \hline\hline
     Redshift range  & $z_{mean}$ & Stellar mass range 	& $\log M_{\star}^{tresh}$		&  $\log M_h^{min}$  	            &  SMHR$\times 10^{2}$ 		& ISFE$\times 10^{2}$ \\ \hline\hline
     $[2.0 - 2.9]$   & $2.50$	& $ [9.10 - 11.40]$    	& $9.10_{-0.16}^{+0.15}$        &  $11.12^{+0.33}_{-0.36}$  & $0.95_{-0.35}^{+0.50}$  	& $6.16_{-2.26}^{+3.23}$  \\ 
     $[2.9 - 5.0]$   & $3.47$	& $ [9.30 - 11.40]$	& $9.30_{-0.19}^{+0.17}$	&  $11.18^{+0.56}_{-0.70}$  & $1.32_{-0.57}^{+0.98}$  	& $8.52_{-3.68}^{+6.32}$  \\
     $[2.0 - 5.0]$\footnote{\tiny high mass sample}   & $3.00$	& $ [9.87 - 11.40]$	& $9.87_{-0.15}^{+0.13}$	&  $11.47^{+0.38}_{-0.43}$  & $2.51_{-0.89}^{+1.23}$  	& $16.19_{-5.74}^{+7.94}$  \\ \hline
 \end{tabular}} \par
  \vspace{-0.75\skip\footins}
  \renewcommand{\footnoterule}{}
 \end{minipage}}
  \label{tab:SHM}
\end{table*}



\section{The VUDS Data}
\label{sec:2_data}
The VIMOS Ultra Deep Survey (VUDS) is a spectroscopic survey of $\sim$10\,000 galaxies performed with the VIMOS multi-object spectrograph at the European Southern Observatory Very Large Telescope \cite{OLF2003}.
Its main aim is to study early phases of galaxy formation and evolution at $2<z<6$.
Details about the survey strategy, target selection, as well as data processing and redshift measurements are presented in ~\cite{OLF2014}. 

We use data in the redshift range $2<z<5$ from two independent fields, COSMOS and VVDS-02h, covering a total area $0.81$ $deg^2$,
corresponding to a volume $\sim$3$\times$10$^7$Mpc$^3$. 
The sample used here contains $3022$ galaxies with reliable spectroscopic redshifts (spectroscopy reliability flags 2, 3, 4 and 9, see \citealt{OLF2014}) and with a stellar mass in the range 9$<$$\log$($M_*$)$<$11 M$_{\sun}$ 
as presented in Figure \ref{fig:stellar_mass}.
The whole sample has been divided into two redshift ranges: $2<z<2.9$ with $\log M_{\star}^{tresh}$=9.1 M$_{\sun}$ and $2.9<z<5.0$ for which $\log M_{\star}^{tresh}$=9.3 M$_{\sun}$, where M$_{\star}^{tresh}$ is the lower mass boundary of the sample resulting from the survey selection function (see below).
Additionally, to estimate the SMHR for more massive galaxies we define a galaxy sub-sample in the range $2<z<5$ and with $\log M_{\star} > 9.87 M_{\sun}$.
This mass limit is the practical limit for which we can measure a galaxy correlation function signal accurately enough at each observed scale 0.3$<r_p<$17 $h^{-1}$Mpc, which is required in order to get the HOD fit to converge.

\section{M$_{\star}$ and M$_h$ measurements}
\label{sec:3_methods}
The stellar masses in the VUDS survey are estimated by performing SED fitting on the multi-wavelength photometry using the 'Le Phare' code (\citeauthor{Ilbert2006} \citeyear{Ilbert2006}), as described in details by \cite{Ilbert2013} and references therein.

Halo masses M$_h$ are measured from a two-step process.
First, the projected two-point correlation function $w_p(r_p)$ is computed for all three sub-samples in \cite{Durkalec2014}. 
The correlation function results are then interpreted in terms of a three-parameter halo occupation model (HOD)
of the form proposed by \cite{Zehavi2005} and motivated by \cite{Kravtsov2004}, with the mean number of galaxies:

\begin{equation}
\label{eq:hodmodel}
\langle N_g|M\rangle = \left\{ \begin{array}{ll}
1 + \left(\frac{M}{M_1}\right)^{\alpha}\quad \textrm{for} & M > M_{min}\\
0 & \textrm{otherwise,}\\
\end{array} \right.
\end{equation}
where $M_{min}$ is the minimum mass needed for a halo to host one central galaxy, and $M_1$ is the mass of a halo having on average one satellite galaxy, while $\alpha$ is the power law slope of the satellite mean occupation function.

The correlation function measurements and model fitting procedures are described in \cite{Durkalec2014}. 
By construction of the halo occupation function given in Eq. \ref{eq:hodmodel}, the parameter $M_{min}$ is the halo mass associated to galaxies with
a stellar mass defined as the stellar mass threshold in the SHM relation \citep{Zheng05,Zehavi2005}. 
We therefore quote the lowest mass of the sample considered as $M_{\star}^{tresh}$, as imposed by the survey limiting magnitude. 
The errors associated to this lower limit have been computed as the average of the errors on M$_{\star}$ from the SED fitting for each redshift and mass sub-sample separately. 

\section{The stellar mass -- halo mass relation at z$\sim$3}
\label{sec:4_results}

\begin{figure*}[t!]
\centering
\begin{tabular}{ccc}
 \hspace*{-0.8cm}\includegraphics[height=8.8cm]{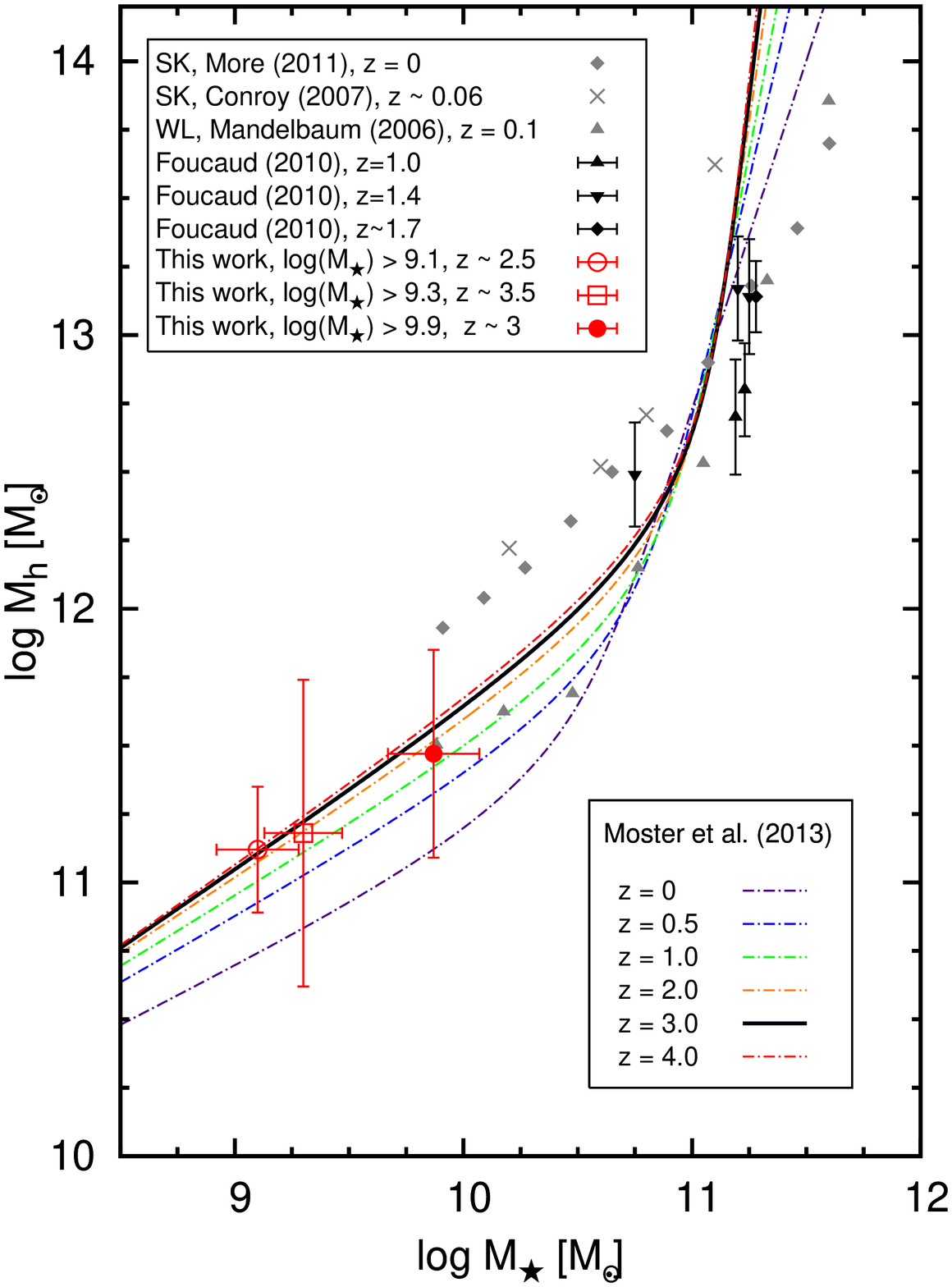} & \hspace*{-0.4cm}\includegraphics[height=8.8cm]{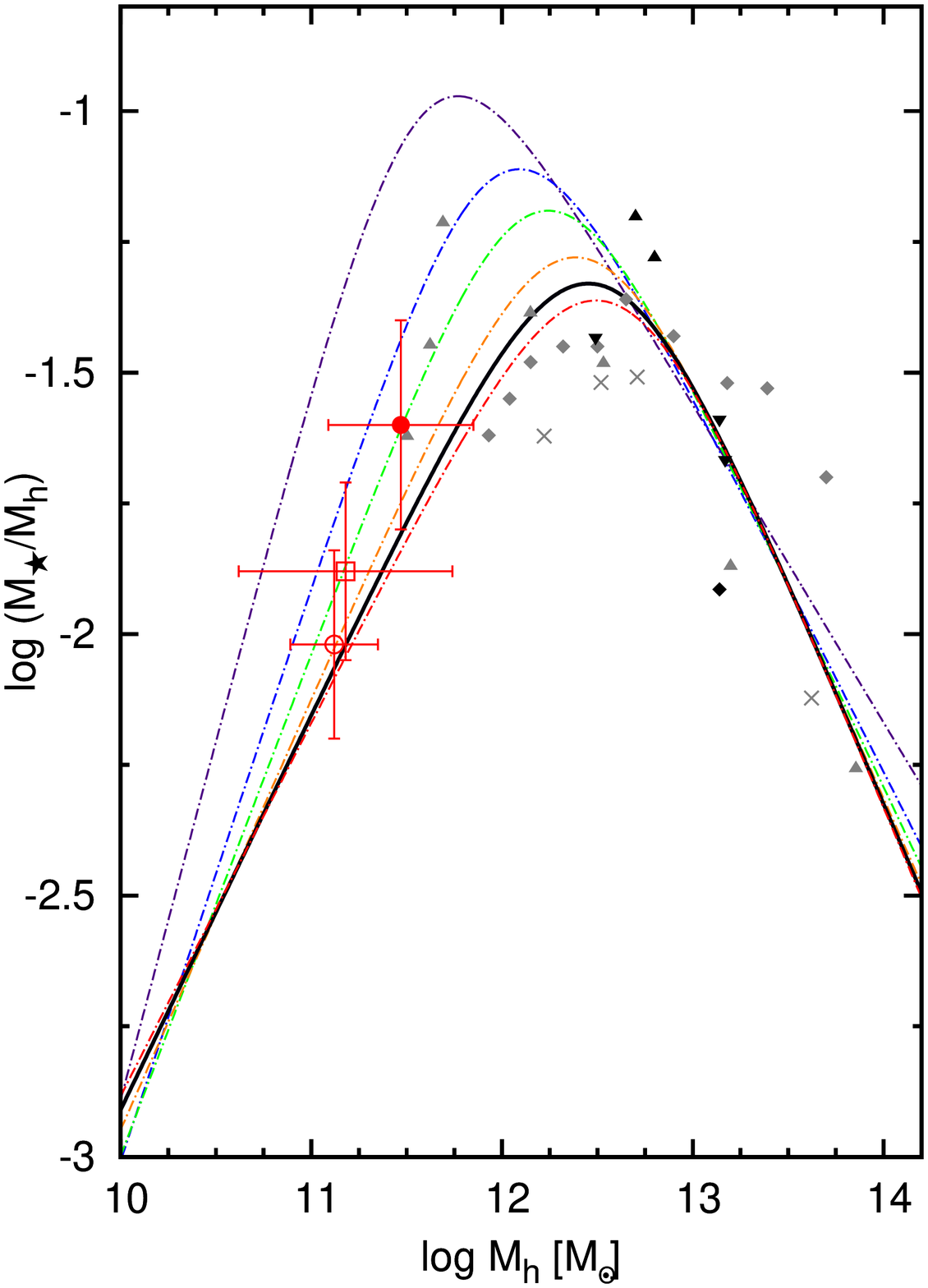} &\hspace*{-0.4cm}\includegraphics[height=8.8cm]{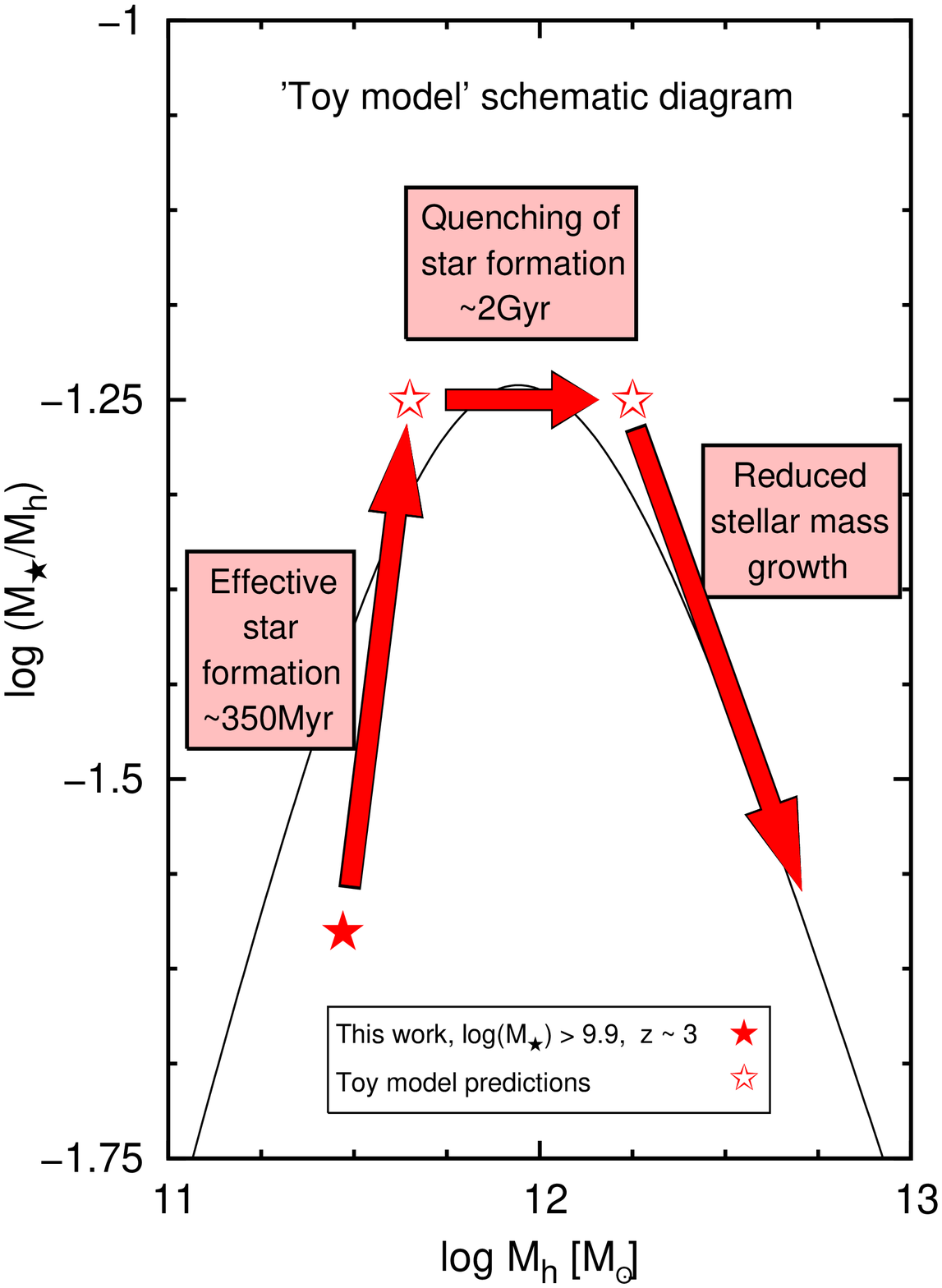} \\
\end{tabular}
 \caption{\textit{Left:} The relation between the stellar mass M$_{\star}$ and the halo mass M$_h$ in  VUDS  for different M$_{\star}$ and redshifts (red symbols).  
   M$_{\star}$ is derived from SED fitting of the multi-wavelength photometric data using known spectroscopic redshifts; 
   error bars in M$_{\star}$ indicate expected uncertainties of the SED fitting method. 
   M$_h$ is obtained from HOD modelling of the two-point correlation function in different redshift and mass ranges.  
   The VUDS data is compared to low and intermediate redshift measurements from satellite kinematics \citep{Conroy2007,More2011}
   weak lensing \citep{Mandelbaum2006}, galaxy clustering \citep{Foucaud2010}.
   The lines represents model predictions derived from abundance matching at various redshift \citep{Moster2013}.
   \textit{Center:} The stellar mass $M_{\star}$ over halo mass $M_{h}$ ratio
   vs. halo mass at $z=3$ in the VUDS survey. The colour scheme is the same as for the left panel.
   \textit{Right:}  Evolution of the $M_{\star}/M_h$ ratio with time predicted from stellar and halo mass accretion histories for the most massive galaxy population observed at $z\sim3$, using the 'toy model' described in the text.}
\label{fig:SHMR}
\end{figure*}

Our results are presented in Table \ref{tab:SHM} and in the left panel of Figure \ref{fig:SHMR}.
We find that for the low redshift sample z$\sim$2.5 the stellar mass for halos of mass $\log M_h^{min}$=$11.12\pm0.33 M_{\sun}$ is $\log M_{\star}^{tresh}$=$9.1 M_{\sun}$, while at z$\sim$3.5 the halo mass reaches $\log M_h^{min}$=$11.18\pm0.56 M_{\sun}$ for a stellar mass $\log M_{\star}^{tresh}$=$9.3 M_{\sun}$. 

From these measurements we find that $\log(M_{\star}/M_h)$ is ranging from $-2.02\pm0.18$ for the low mass sample up to
$-1.6\pm0.17$ for the most massive sample, at a redshift z$\sim$3. 
As shown in Figure \ref{fig:SHMR} these results are compared to various measurements at low and intermediate redshift z$<$1,
obtained using different methods, including satellite kinematics (\citeauthor{Conroy2007} \citeyear{Conroy2007}, \citeauthor{More2011} \citeyear{More2011}), weak lensing (\citeauthor{Mandelbaum2006} \citeyear{Mandelbaum2006}), galaxy clustering (\citeauthor{Foucaud2010} \citeyear{Foucaud2010}, \citeauthor{Hartley2013} \citeyear{Hartley2013}),  
as well as abundance matching \citep{Moster2013}.
Our measurements are in excellent agreement with models derived from abundance matching at a redshift z$=$3 \citep{Moster2013}.

\section{Discussion and conclusion}
\label{sec:5_discussion}

Our SMHR measurements are among the first performed at z$\sim$3 from a clustering
and HOD analysis, as made possible from the large VUDS spectroscopic redshift survey. 
We find that the SMHR is 1\% to 2.5\% for galaxies with intermediate stellar masses 
(at z$\sim$3) ranging from $\sim$10$^9$ M$_{\sun}$ to $\sim$7$\times10^9$ M$_{\sun}$.

Following \cite{Conroy2009a} we compute the integrated star formation efficiency (ISFE)
$\eta$=M$_{\star}/$M$_h / f_b$ with $f_b$ the universal baryon fraction   $f_b=\Omega_b / \Omega_m=0.155$ (Planck collaboration 2014).
Results are reported in Table \ref{tab:SHM}.
We find that the ISFE range from 6.2$^{+3.2}_{-2.3}$\% to 16.2$^{+7.9}_{-5.7}$\% for galaxies with M$_{\star}$ from M$_{\star}^{tresh}$=$1.3\times10^9$ $M_{\sun}$ 
to M$_{\star}^{tresh}$=$7.4\times10^9$ $M_{\sun}$. The IFSE at z$\sim$3 therefore increases with M$_{\star}$ over this mass range. 
The star formation efficiency of $\sim$16\% in a halo with M$_h$=$3\times10^{11}$ M$_{\sun}$ is quite close to a 
maximum of $\sim$20\% occurring at $10^{12}$ M$_{\sun}$ in SMHM models \cite{Behroozi2010} (see also \citealt{Moster2013}).

We use a simple mass growth model to derive the time scale for which our most massive galaxy sample would reach the maximum predicted in the SMHM relation from \cite{Behroozi2010}.  
In this model the mass growth of DM haloes is described by a mean accretion rate $\langle\dot{M_H} \rangle_{mean}$ taken from \cite{Fakhouri2010}, while galaxies grow in M$_{\star}$ via star formation using the median SFR for our sample \citep{Tasca2014}, as well as through mergers with a constant accretion in stars of $\sim 1 M_{\sun}/yr$ \citep{Tasca2014b}.
We compute the halo and stellar mass values every $\delta t=$ 5 Myr to account for the halo accretion rate and SFR changing with redshift and mass. In the right panel of Figure \ref{fig:SHMR} we represent the expected time evolution in $M_{\star}/M_h$ versus M$_{\star}$ for a galaxy starting at z$\sim$3  following this 'toy model'.
We find that the SMHM relation would reach a maximum $\log(M_{\star}/M_h)\simeq-1.25$ about 360Myr after the observed epoch (i.e. at $z\sim2.6$) and that at this time halo and stellar masses will be $M_h^{min}=10^{11.6}M_{\sun}$ and $M_{\star}^{trash}$=2$\times 10^{10}M_{\sun}$ respectively.

According to the model proposed by \cite{Moster2013} the SMHM relation should turn over after reaching a maximum, with the slope of the relation maintaining the same absolute value but reversing sign (see Figure \ref{fig:SHMR}).
Since dark matter halos grow in time \citep[e.g.][]{Fakhouri2010} the growth in stellar mass must drop dramatically over a sustained period of time in order to follow a change in $M_{\star}/M_h$ by roughly an order of magnitude in the SMHM, or, alternatively, the dark matter accretion rate $\langle \dot{M_{H}} \rangle$ must rise precipitously, or a combination of the two.
There are no indications e.g. in N-body simulations which support a dramatic sustained rise in $\langle \dot{M_{H}}\rangle$.
On the other hand the stellar mass computed at the maximum of the SMHM relation is $M_{\star}\sim2\times10^{10}$ $M_{\sun}$,  comparable to the ’quenching mass’ as discussed in \cite{Bundy2006} and is massive enough at z$\sim$2.6 that mass-related quenching may be dominant \citep{Peng2010}. In this picture, on average, the massive galaxy population in VUDS with $M_{\star}^{min}$$\sim$0.5-1$\times10^{10}$ M$_{\sun}$ observed at $z\sim3$ will experience star formation quenching within a few hundred million years. 
Assuming that the SFR will successfully drop and will reach zero when the halo mass grows by 0.5 dex (the approximate width the peak of the SMHM relation), we compute that the star formation will then be quenched within $\sim2$Gyr (by $z\sim1.5$).
After that time the stellar mass would grow only slowly e.g. through mergers and/or lower levels of star formation in order for galaxy populations to follow the SMHM relation.


In conclusion, the SMHM is a simple yet efficient tool to probe star formation efficiency at the epoch of rapid 
stellar mass assembly provided one obtains robust measurements on both M$_{\star}$ and M$_h$; this is now possible
with VUDS at z$\sim$3, complementing more indirect estimates using e.g. abundance matching. 
A more extensive exploration of the efficiency of star formation over a larger range of halo
masses is becoming possible with new surveys, and it would be interesting to
probe higher masses than done in this paper to evaluate the halo mass corresponding to the 
highest star formation efficiency. Extending such measurements to higher redshifts
will require the power of new facilities like PFS-Sumire, JWST or ELTs.  



\begin{acknowledgements}
We thank Jean Coupon and Carlo Schimd for interesting discussions.
This work is supported by the European Research Council
Advanced Grant ERC-2010-AdG-268107-EARLY, and by INAF Grants PRIN 2010\&2012 and PICS 2013. 
AC, OC, MT and VS acknowledge the grant MIUR PRIN 2010--2011.  
This work is supported by the OCEVU Labex (ANR-11-LABX-0060) and the A*MIDEX project (ANR-11-IDEX-0001-02).
AP is supported by grant UMO-2012/07/B/ST9/04425 and the Polish-Swiss Astro Project. 
Research conducted within the scope of the HECOLS International Associated Laboratory, supported in part by the Polish NCN grant DEC-2013/08/M/ST9/00664.
This work is based on data products made available
at the CESAM data center, Laboratoire d'Astrophysique de Marseille, France.
\end{acknowledgements}

\bibliographystyle{aa}
\bibliography{shmr}

\end{document}